\documentclass[a4paper,11pt]{article}
\pdfoutput=1 

\usepackage{jcappub} 

\usepackage[T1]{fontenc} 

\title{The effect of dark energy on the void-halo perpendicular alignments}

\author{Geonwoo Kang,}
\author[1]{Jounghun Lee \note{Corresponding author.}}

\affiliation{Department of Physics and Astronomy, Seoul National University, \\
Kwanak-ro 1, Kwanak-gu, Seoul 08826, Republic of Korea}

\emailAdd{kanggeonwoo@snu.ac.kr}
\emailAdd{cosmos.hun@gmail.com}

\abstract{We report a numerical discovery that in a more rapidly accelerating spacetime, the galactic halos on void surfaces develop stronger perpendicular alignments 
with the directions toward the void centers. 
We utilize the halo catalogs from the AbacusSummit suite of simulations for $10$ different cosmologies that include one Planck $\Lambda$CDM, four $w$CDM and 
five $w_{0}w_{a}$CDM, which share the identical initial conditions except for the dark energy equation of state. 
For each cosmology, we identify the voids via the Void-Finder algorithm~\cite{HV02} at $z=0.1$ and look for the void-surface galactic halos in the mass range of 
$11.5\le \log M/(h^{-1}\,M_{\odot})< 13$. Then, we determine the probability density functions of the cosines of the angles, $p(\cos\theta)$, between the shape axes 
of void-surface galactic halos and the directions toward the void centers. 
The numerically obtained  $p(\cos\theta)$ is fitted to an analytic single-parameter formula derived through an empirical modification of the linear perturbation theory~\cite{KL26}. 
Eliminating spurious signals caused by the differences in the mass and sphericity distributions of void-surface galactic halos among different cosmologies,  
we detect a clear net effect of dark energy on the strengths of the perpendicular alignments of void-surface galactic halos, quantified by the single parameter, $d_{t}$. 
Noting that $d_{t}$ has higher values in the cosmologies where dark energy has more negative pressure and evolves more rapidly, we put forth a bilinear model 
for the difference in $d_{t}$ between the two cases of $w=-1$ and $w\ne -1$: $\Delta d_{t}(w_{0},w_{a}) = \alpha(1+\,w_{0})+\beta\,w_{a}$, 
with two universal coefficients, $\alpha$ and $\beta$. Demonstrating that this bilinear relation excellently describes the numerical results, we conclude that the perpendicular 
alignments of void-surface galactic halos should in principle be a powerful independent indicator of the dynamic nature of dark energy.}
\begin{document} 
\maketitle
\flushbottom

\section{Introduction}\label{sec:intro}

The void-galaxy perpendicular alignment designates a phenomenon that the giant galaxies located on void surfaces with stellar masses $\log\left[M_{\star}/(h^{-1}M_{\odot})\right]\ge 10.5$ 
exhibit a strong tendency of having their shape axes lie on the tangential planes parallel to the void surfaces and normal to the directions toward the void centers. The occurrence of this 
phenomenon was recently reported in the study of ref.~\cite{KL26} based on the IllustrisTNG 300-1 hydrodynamic simulations~\cite{tng1,tng2,tng3,tng4,tng5,tng6} 
and was found to be a unique feature of the void-surface galaxies, as no significant signals of perpendicular alignments were detected from the galaxies outside and inside the void boundaries. 
In the same study was found that the stronger perpendicular alignments were exhibited by those void-surface galaxies having faster rotational motions, earlier formation epochs, 
redder colors and lower specific star formation rates~\cite{KL26}. These numerical results were interpreted as evidence supporting the scenario that the void-surface galaxies develop stronger 
perpendicular alignments if they are exposed for a longer period of time to the rapid expansion of cosmic voids, which effectively induces tangential flow of matter and gas particles on void 
surfaces. 

In this scenario, it is naturally speculated that the strengths of void-galaxy perpendicular alignments should depend on the amount and equation of state of dark energy (DE). 
A larger amount  and stronger anti-gravity of DE would lead the cosmic voids to form earlier, which in turn would enhance the perpendicular alignments of void-surface galaxies as 
they should be exposed to the effect of void expansions for a longer period of time. Especially, in a universe with DE characterized by a more negative equation of state like the 
phantom models~\cite{phantom_review}, the void formation will also accelerate ever more, leading the void-surface galactic halos to acquire stronger perpendicular alignments 
with the directions toward the void centers. If this speculation turns out to be the case, then the perpendicular alignments of void-surface galaxies could be a new non-linear statistics 
that can be used to put an independent and complementary constraint on the DE equation of state. 

In this paper, we attempt to validate this speculation by utilizing N-body simulations performed for various DE cosmologies. We intend to numerically explore if and how the strengths of 
perpendicular alignments of void-surface galaxies vary with the DE equation of state, $w$, when the other key cosmological parameters remain unchanged from the values of the standard 
cosmology.  In the current work will be considered three different DE candidates: the standard cosmological constant $\Lambda$ whose equation of state is a perfect constant 
of $w=-1$, the quintessence scalar field having a constant equation of state, $w\ne -1$~\cite{RP88,wet88}, and the dynamical scalar-field DE with time varying equation of state, 
$w(a)=w_{0}+(1-a)w_{a}$ where $a$ is the scale factor and two coefficients are defined as $w_{0}\equiv w(a=1)$ and $w_{a}\equiv -dw(a)/da$~\citep{CP01,linder03}.
In all of the cosmologies considered, two assumptions are made: the most dominant matter contents are the cold dark matter (CDM) and the spatial curvature is zero (i.e., flat geometry). 
Following the convention, the cosmologies with a flat geometry, CDM and these three different DE will be referred to as the $\Lambda$CDM, $w$CDM and 
$w_{0}w_{a}$CDM, respectively, throughout this paper.  
 
In section~\ref{sec:review} is presented a brief review on the analytic derivation of the single-parameter formula for the perpendicular alignments of void-surface galaxies.
In section~\ref{sec:lcdm} is presented the comparison between the numerical results on the perpendicular alignments of void-surface galaxies and the analytic formula with 
best-fit single parameter for the $\Lambda$CDM cases, and demonstrated the variation of the best-fit single parameter with the initial conditions.  
In section~\ref{sec:wcdm} is presented the same as in section~\ref{sec:lcdm} but for the $w$CDM and $w_{0}w_{a}$CDM cosmologies and displayed the dependence of  
the single parameter on the DE equation of state. In section~\ref{sec:con} is discussed the upside and downside of this new probe of DE and drawn a 
final conclusion.

\section{Perpendicular alignments of void-surface galactic halos}

\subsection{A concise review of the analytic model}\label{sec:review}

The derivation of the analytic formula for the perpendicular alignments of void-surface galactic halos~\cite{KL26} was based on the theoretical framework established 
by ref.\cite{lee19} to describe the intrinsic shape alignments of DM halos with the principal axes of surrounding tidal fields, ${\bf T}=(T_{ij})$. 
This framework starts with the key assumption that the eigenvector, ${\bf e}$, of the halo inertia tensor corresponding to the largest eigenvalue follows a 
multi-variate Gaussian probability distribution, $p({\bf e})$, characterized by the following covariance matrix, ${\bf \Sigma}=(\Sigma_{ij})$:
\begin{equation}
\label{eqn:eiej}
\Sigma_{ij} = \frac{1+d_{t}}{3}\delta_{ij} - d_{t}\hat{T}_{ij}\, ,\,\,\, {\rm for}\,\,\, i,j\in\{1,2,3\}\, ,
\end{equation}
where $\hat{\bf T}$ is the unit traceless version of ${\bf T}$ satisfying $\vert\hat{\bf T}\vert=1$ and ${\rm Tr}(\hat{\bf T})=0$, and $d_{t}\in [0,1)$ is a free parameter introduced 
to effectively quantify the strengths of the alignments of ${\bf e}$ with the principal axes of ${\bf T}$.  

In the local spherical-polar coordinate system, the marginalization of $p({\bf e})$ over the magnitude of ${\bf e}$, i.e., $e\equiv \vert{\bf e}\vert$, 
yields the probability density function of the unit vector,  $\hat{\bf e}\equiv {\bf e}/e$, which will be referred to as the halo shape axis from here on:
\begin{equation}
\label{eqn:phu}
p(\hat{\bf e})=\int_{0}^{\infty} p(e\hat{\bf e})e^{2}de=\frac{1}{4\pi{\rm det}({\bf\Sigma})^{1/2}}\left(\hat{\bf e}\cdot{\bf {\Sigma}}^{-1}\cdot\hat{\bf e}\right)^{-3/2}\, ,
\end{equation}
where $\textrm{det}\!\left({\bf \Sigma}\right)$ is the determinant of ${\bf \Sigma}$. 
If the orthonormal basis vectors, $\{\hat{\bf r},\hat{\bf \theta},\hat{\bf \phi}\}$, of the local spherical polar-coordinates are chosen to be parallel to three principal axes 
of $\hat{\bf T}$, eq.~(\ref{eqn:phu}) can be expressed as
\begin{eqnarray}
\label{eqn:phu_prin}
p(\hat{\bf e}) 
&=& \frac{1}{2\pi}\left[\prod_{i=1}^{3}\left(1+d_{t}-3d_{t}\hat{\lambda}_{i}\right)\right]^{-\frac{1}{2}}
\left(\sum_{i=1}^{3}\frac{\vert\hat{\bf t}_{i}\cdot\hat{\bf e}\vert^{2}}{1+d_{t}-3d_{t}\hat{\lambda}_{i}} \right)^{-\frac{3}{2}}\, ,
\end{eqnarray}
where $\{\hat{\bf t}_{1},\hat{\bf t}_{2},\hat{\bf t}_{3}\}$ are three eigenvectors of $\hat{\bf T}$ corresponding to its three eigenvalues, $\{\hat{\lambda}_{1},\hat{\lambda}_{2},\hat{\lambda}_{3}\}$, 
in a decreasing order, $\hat{\lambda}_{1}\ge \hat{\lambda}_{2}\ge\hat{\lambda}_{3}$, satisfying $\sum_{i=1}^{3}\hat{\lambda}_{i}=0$ and $\sum_{i=1}^{3}\hat{\lambda}^{2}_{i}=1$. 

The validity of eq.~(\ref{eqn:phu_prin}) is guaranteed provided that the tidal field is smoothed on the large scales where the approximation of ${\bf T}$ as a Gaussian random 
field holds true and that the DM halos retain well the dominant influences from the surrounding tidal fields even after their relaxations.  
Given these prerequisites, the optimal targets for an application of eq.~(\ref{eqn:phu_prin}) should be the void-surface halos, being embedded in the mildly dense environments where 
the secondary nonlinear effects on the halo shapes other than the tidal influences are not so strong and surrounded by the large-scale tidal fields smoothed on the scales of void radii. 
Furthermore, for the cases of void-surface halos, the major principal axes of ${\bf T}$ can be readily determined without reconstructing the full tidal fields as the directions from the halo 
positions toward the void centers~\cite{KL26}.

Focusing on the alignments between the shape axes of void-surface galactic halos and the directions toward the void centers which are parallel to the major principal axes of the tidal 
fields, we orient the orthonormal base vectors, $\{\hat{\bf r},\hat{\bf \theta},\hat{\bf \phi}\}$, to be parallel to the {\it major}, {\it intermediate}, and {\it minor} principal axes 
of $\hat{\bf T}$, i.e., $\hat{\bf r}=\hat{\bf t}_{1}$, $\hat{\bf \theta}=\hat{\bf t}_{2}$, and $\hat{\bf \phi}=\hat{\bf t}_{3}$. Then, we marginalize $p(\hat{\bf e})$ over the azimuthal angle 
$\phi$ to obtain the probability density function of the cosines of the polar angles, $p(\cos\theta)$: 
\begin{eqnarray}
p(\cos\theta)
&=& \frac{1}{2\pi}\left[\left(1+d_{t}\right)\left(1+d_{t}-3d_{t}/\sqrt{2}\right)\left(1+d_{t}+3d_{t}/\sqrt{2}\right)\right]^{-\frac{1}{2}}\times \nonumber \\
\label{eqn:pcost}
&& 
\int_{0}^{2\pi}d\phi\left[\frac{\cos^{2}\theta}{1+d_{t}-3d_{t}/\sqrt{2}} + \frac{(1-\cos^{2}\theta)\cos^{2}\phi}{1+d_{t}} + \frac{(1-\cos^{2}\theta)\sin^{2}\phi}{1+d_{t}+3d_{t}/\sqrt{2}}\right]^{-\frac{3}{2}}\, ,
\end{eqnarray}
where $\cos\theta\equiv \vert\hat{\bf e}\cdot\hat{\bf t}_{1}\vert$, $\sqrt{1-\cos^{2}\theta}\cos\phi = \vert\hat{\bf e}\cdot\hat{\bf t}_{2}\vert$ and  
$\sqrt{1-\cos^{2}\theta}\sin\phi = \vert\hat{\bf e}\cdot\hat{\bf t}_{3}\vert$. Here, the three eigenvalues of $\hat{\bf T}$ are 
set at the approximate values of $\hat{\lambda}_{1}=1/\sqrt{2}$, $\hat{\lambda}_{2}=0$ and $\hat{\lambda}_{3}=-1/\sqrt{2}$. 
If $d_{t}>0$ (corresponding to the perpendicular alignments between $\hat{\bf e}$ and $\hat{\bf t}_{1}$), then eq.(\ref{eqn:pcost}) predicts the decrease of $p(\cos\theta)$ 
with the increment of $\cos\theta$.  A larger positive value of $d_{t}$ indicates stronger perpendicular alignments between the shape axes of void-surface halos and the 
directions toward the void centers. 

\subsection{Numerical results for the $\Lambda$CDM cases}\label{sec:lcdm}
\begin{table*}[tbp]
\centering
\begin{tabular}{lcccccc}
\hline
\hline
\rule{0pt}{3.5ex}\noindent
Model & $\Omega_{m}$ & $\sigma_8$ & $n_s$ & $N_{\rm void}$ & $N_{\rm vs}$ & $10^2 d_t$ \\
\hline
\rule{0pt}{3.5ex}\noindent
c000 & $0.3138$ & $0.8114$ & $0.9649$  & 634051 & 41432782 & $7.80 \pm 0.02$ \\
\hline
\rule{0pt}{3.5ex}\noindent
  c004 & $0.3138$ & $0.7532$ & $0.9649$  & 633030 & 41277401 & $8.31 \pm 0.02$ \\
  c112 & $0.3138$ & $0.8276$ & $0.9649$  & 634335 & 41467584 & $7.68 \pm 0.03$\\
  c113 & $0.3138$ & $0.7954$ & $0.9649$ & 633647 & 41409982  & $7.94 \pm 0.03$ \\
  c116 & $0.3138$ & $0.8698$ & $0.9649$ & 636013 & 41502889 & $7.37 \pm 0.03$\\
  c125 & $0.3138$ & $0.8154$ & $0.9649$  & 634266 & 41442900 & $7.77 \pm 0.02$\\
  c126 & $0.3138$ & $0.8073$ & $0.9649$ & 633930 & 41417703 & $7.81 \pm 0.02$  \\
  c130 & $0.3138$ & $0.7142$ & $0.9649$ & 633420 & 41129776 & $8.70 \pm 0.03$ \\
  c133 & $0.3138$ & $0.9090$ & $0.9649$ & 637751 & 41500517 & $7.10 \pm 0.05$ \\ 
\hline
\rule{0pt}{3.5ex}\noindent
  c102 & $0.3363$ & $0.8116$ & $0.9649$ & 616537 & 41735888 & $7.71 \pm 0.03$ \\
  c103 & $0.2928$ & $0.8116$ & $0.9649$ & 649173 & 40934517 & $7.90 \pm 0.04$\\
  c131 & $0.2551$ & $0.8116$ & $0.9649$ & 672298 & 39406575 & $8.09 \pm 0.03$\\
  c134 & $0.3881$ & $0.8116$ & $0.9649$  & 572645 & 41625173 & $7.45 \pm 0.03$ \\
  \hline
\rule{0pt}{3.5ex}\noindent
 c104 & $0.3138$ & $0.8116$ & $0.9749$  & 630735 & 41258217 & $7.76 \pm 0.04$\\
 c105 & $0.3138$ & $0.8116$ & $0.9549$  & 636963 & 41603818 & $7.86 \pm 0.03$\\
\hline
\end{tabular}
\caption{\label{tab:lcdm}
Appellation of the background $\Lambda$CDM cosmologies, three key cosmological parameters, total numbers of voids and void-surface halos, 
and best-fit single parameter. For all of the cosmologies listed, the $\Lambda$ density parameter is given as $\Omega_{\Lambda}=1-\Omega_{m}$, and 
the dimensionless Hubble parameter $h$ is set at the value that meets the cosmic microwave background constraints~\cite{summit1}. }
\end{table*}

Numerical datasets utilized for our analysis are all from the AbacusSummit~\cite{summit1,summit2} suite of DM only $N$-body simulations conducted with the Abacus code \citep{summit_code} 
for a broad range of background cosmologies. 
Among several different volume runs of the AbacusSummit simulations, we exclusively rely on the volume run of $8\,h^{-3}\,{\rm Gpc}^{3}$ and consider three different classes of the background models: 
the standard $\Lambda$CDM, $w$CDM and $w_{0}w_{a}$CDM cosmologies, the initial conditions of which were all generated by the {\it Cosmic Linear Anisotropy Solving System} (CLASS)~\cite{class}.
In the simulations are involved a total of $6912^{3}$ CDM particles with individual mass $2\times 10^{9}\,h^{-1}\,M_{\odot}$, the gravitational 
forces among which were computed according to the analytical split method developed by M.~V.~L.~Metchnik~\cite{met09} from the initial redshift $z=99$ down to 
$z=0.1$. To the DM particle snapshot of each AbacusSummit simulation at a given redshift was applied the COMPASO halo-finding 
scheme~\cite{summithalo1,summithalo2}, an upgraded version of the classical spherical overdensity (SO) halo-finder~\cite{SO}, to identify gravitationally 
bound objects. 

We first analyze the halo catalog from the AbacusSummit simulation performed for the Planck flat $\Lambda$CDM cosmology~\cite{planck18} (see the second row of table~\ref{tab:lcdm}). 
Applying the Void-Finder algorithm~\cite{HV02} to the halo catalog at $z=0.1$ (the latest epoch considered in the AbacusSummit suite), we follow the exactly same steps as fully 
described in ref.~\cite{KL25} to identify cosmic voids and measure their effective radii, $R_{\rm eff}$. 
For each of the galactic halos in the logarithmic mass range of $11.5\le m\equiv \log[M_{t}/(h^{-1}M_{\odot})]< 13$, we find its nearest void and measure the distance, $r$, from its position 
to the center of its nearest void. If $0.9\le r/R_{\rm eff}\le 1.1$, then it is identified as the void-surface galactic halos, to be consistent with the scheme of ref.~\cite{KL26}. 
The total number of voids and void-surface galactic halos for the Planck $\Lambda$CDM case are all listed in the second row of table~\ref{tab:lcdm}. 

Using information on the three principal axes of the inertia tensor of each halo provided in the AbacusSummit halo catalog, we determine the 
shape axis, ${\bf e}$, of each void-surface galactic halo as the unit eigenvector of ${\bf I}$ corresponding to the largest-eigenvalue. At each void-surface galactic halo 
position, ${\bf x}$, we also determine the direction toward the void center, ${\bf p}_{c}$, as $\hat{\bf r}\equiv ({\bf x}-{\bf p}_{c})/\vert {\bf x}-{\bf p}_{c}\vert$. 
Then, the cosines of the angles between the shape axes of void-surface galactic halos and directions toward the void centers are computed as 
$\cos\theta \equiv \vert{\bf r}\cdot{\bf e}\vert$.  Finally, the probability density, $p(\cos\theta)$, is numerically obtained as $n_{\rm vs}/(N_{\rm vs}\Delta \cos\theta)$ where 
$n_{\rm vs}$ is the number of the void-surface galactic halos whose values of $\cos\theta$ fall in a short interval, $[\cos\theta,\cos\theta+\Delta\cos\theta]$, 
while $N_{\rm vs}$ is the total number of void-surface galactic halos. 
The uncertainty associated with the determination of $p(\cos\theta)$ in each $\cos\theta$-interval is computed through the Jackknife error analysis. 
We first create eight Jackknife resamples of the voids by dividing the whole simulation volume into eight sub-volumes of equal size and sorting out the voids whose centers are located 
in each of the eight sub-volumes.  Then, we separately determine $p(\cos\theta)$ from each resample and compute its one standard deviation scatter 
among the eight resamples as the Jackknife errors. Throughout this paper, this Jackknife analysis is exclusively used to compute the errors associated with the numerical measurements 
of $p(\cos\theta)$.

\begin{figure}[tbp]
\centering 
\includegraphics[width=0.85\textwidth=0 380 0 200]{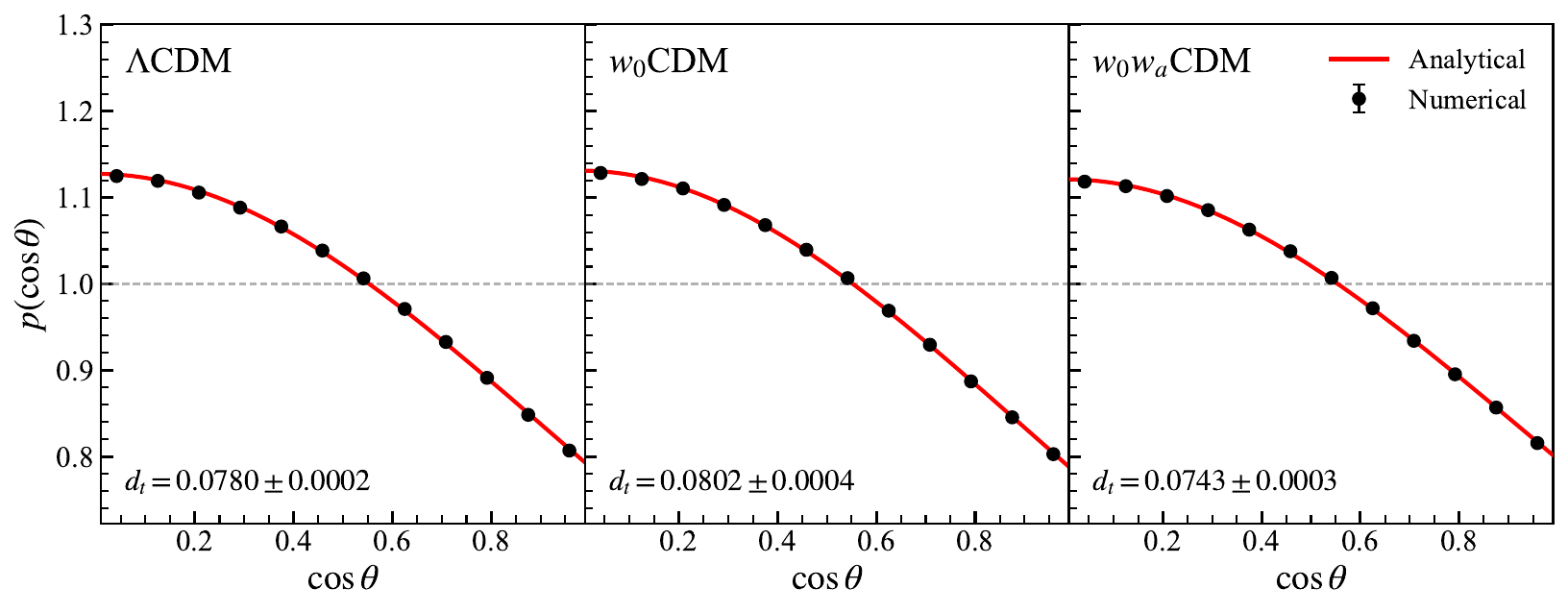}
\caption{\label{fig:pcost} Numerically obtained probability density functions of the cosines of the angles (black filled circles) with Jackknife errors between the shape axes of galactic halos located 
on void surfaces and the directions from the galactic halo positions to the void centers compared with the analytic model given in eq.(\ref{eqn:pcost}) with best-fit coefficient $d_{t}$ for three 
different DE models.}
\end{figure}

Comparing this numerical result to the analytic formula, eq.~(\ref{eqn:pcost}), via the $\chi^{2}$-statistics, we determine the best-fit value of $d_{t}$ that minimizes $\chi^{2}$ 
and find the error in $d_{t}$ from the Gaussian maximum likelihood distribution of $p(-\chi^{2}/2)$.  
The left-panel of figure~\ref{fig:pcost} plots the numerical result of $p(\cos\theta)$ (black circles) with Jackknife errors for the Planck $\Lambda$CDM case at $z=0.1$ and compares it with the 
analytic formula (red solid lines) with the best-fit value of $d_{t}$ (see the fifth column of the second row in Table~\ref{tab:lcdm}). As can be seen, the analytic formula describes very well 
the numerically obtained $p(\cos\theta)$, confirming the validity of eq.~(\ref{eqn:pcost}) for the Planck $\Lambda$CDM case. The shape axes of void-surface galactic halos exhibit strong 
tendency to be perpendicularly aligned with the directions toward the void centers, which is consistent with the results of our prior study based on hydrodynamic simulations where 
the shape axes of void-galactic halos were measured from the stellar particles. This consistency implies that no matter which components of void-surface galactic halos between the stellar 
and DM are used to determine the shape axes, there exist significant signals of the perpendicular alignments with the directions toward the void centers.

\begin{figure}[tbp]
\centering 
\includegraphics[width=0.85\textwidth=0 380 0 200]{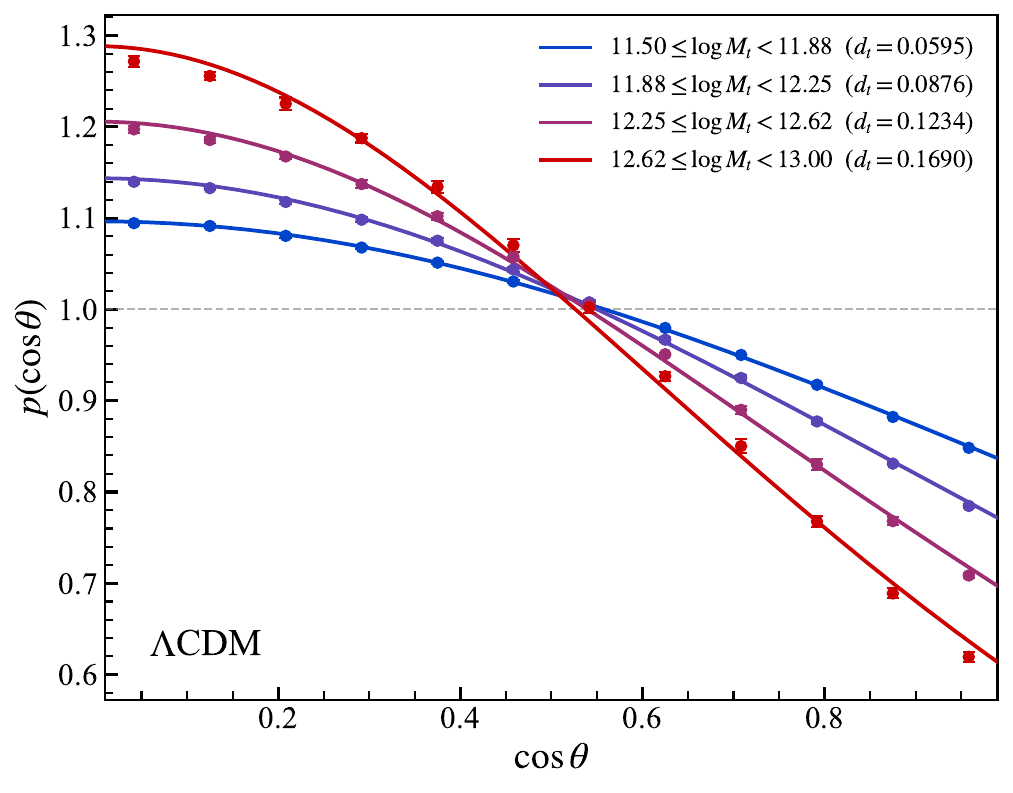}
\caption{\label{fig:pcost_m} 
Differences in the behaviors of $p(\cos\theta)$ among four different mass ranges of void-surface galactic halos for the $\Lambda$CDM case, 
showing a consistent trend that the more massive galactic halos on void surfaces have stronger perpendicular alignment tendencies with the directions 
toward the void centers. The errors in $p(\cos\theta)$ are computed via the Jackknife analysis.}
\end{figure}

\begin{figure}[tbp]
\centering 
\includegraphics[width=0.85\textwidth=0 380 0 200]{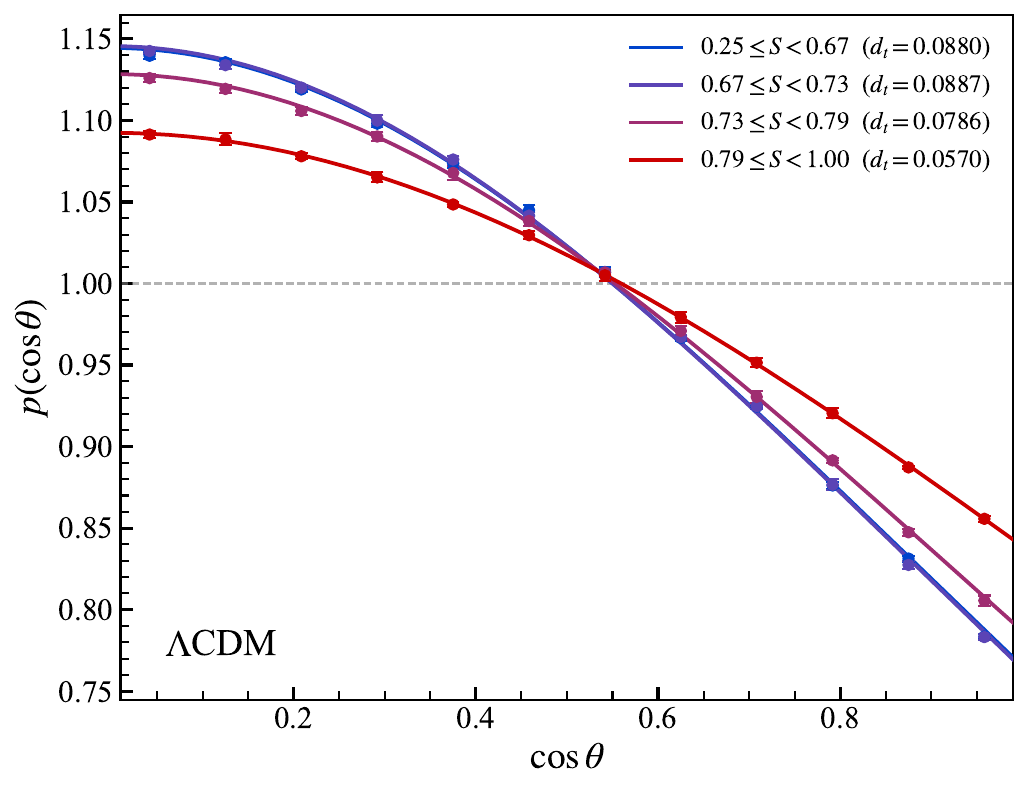}
\caption{\label{fig:pcost_s} 
Differences in the behaviors of $p(\cos\theta)$ among four different sphericity ranges of void-surface galactic halos for the $\Lambda$CDM case, 
indicating a consistent trend that the more non-spherical galactic halos on void surfaces have stronger perpendicular alignment tendencies with the directions 
toward the void centers. The errors in $p(\cos\theta)$ are computed via the Jackknife analysis.}
\end{figure}

Speculating that the strengths of the perpendicular alignments of void-surface galactic halos may depend on the halo mass, $M_{t}$, we split the sample of void-surface galactic halos into four 
subsamples of comparable size according to their logarithmic masses, $m_{t}$, and separately determine $p(\cos\theta)$ from each subsample. 
Figure~\ref{fig:pcost_m} plots the numerical results of $p(\cos\theta)$ (black circles) with Jackknife errors from the four $m_{t}$-selected subsamples for the Planck $\Lambda$CDM case at 
$z=0.1$ and compares them with the best-fit formula (red solid lines). An obvious trend is witnessed: the more massive void-surface galactic halos exhibit stronger perpendicular alignments 
with the directions toward the void centers. 
We also speculate that $d_{t}$ may depend not only on $m_{t}$ but also on the halo sphericity, $S$, defined as $S\equiv \sqrt{\varrho_{3}/\varrho_{1}}$ where $\varrho_{1}$ and $\varrho_{3}$ 
are the largest and lowest eigenvalues of the halo inertia tensors~\cite{bet-etal07}.  Splitting the sample of void-surface galactic halos into four 
subsamples of comparable size according to their values of $S$, we separately obtain $p(\cos\theta)$ from each $S$-selected subsample and fit it to eq.~(\ref{eqn:pcost}), 
the results of which are all shown in figure~\ref{fig:pcost_s}. As can be seen, the more aspherical void-surface galactic halos (i.e., lower value of $S$) yield 
stronger perpendicular alignments with the directions toward the void centers. 

As stated in ref.~\cite{KL26} and in section \ref{sec:intro} as well, the perpendicular alignments of void-surface halos are facilitated by the maximal matter compression along the 
directions of void expansion.  In this picture, it is naturally expected that stronger perpendicular alignments should be exhibited by those void-surface galactic halos that are exposed 
to the void expansion for a longer period of time.  Therefore, in a universe where the cosmic voids tend to form earlier, the void-surface galactic halos would develop stronger perpendicular 
alignments with the directions toward the void centers. Such a universe would be characterized by a larger amount of DE (higher $\Omega_{\Lambda}$), a lower amplitude of initial density power 
spectrum (lower $\sigma_{8}$) and a milder slope on the super-horizon scale (lower spectral index $n_{s}$). 


\begin{figure}[tbp]
\centering 
\includegraphics[width=0.85\textwidth=0 380 0 200]{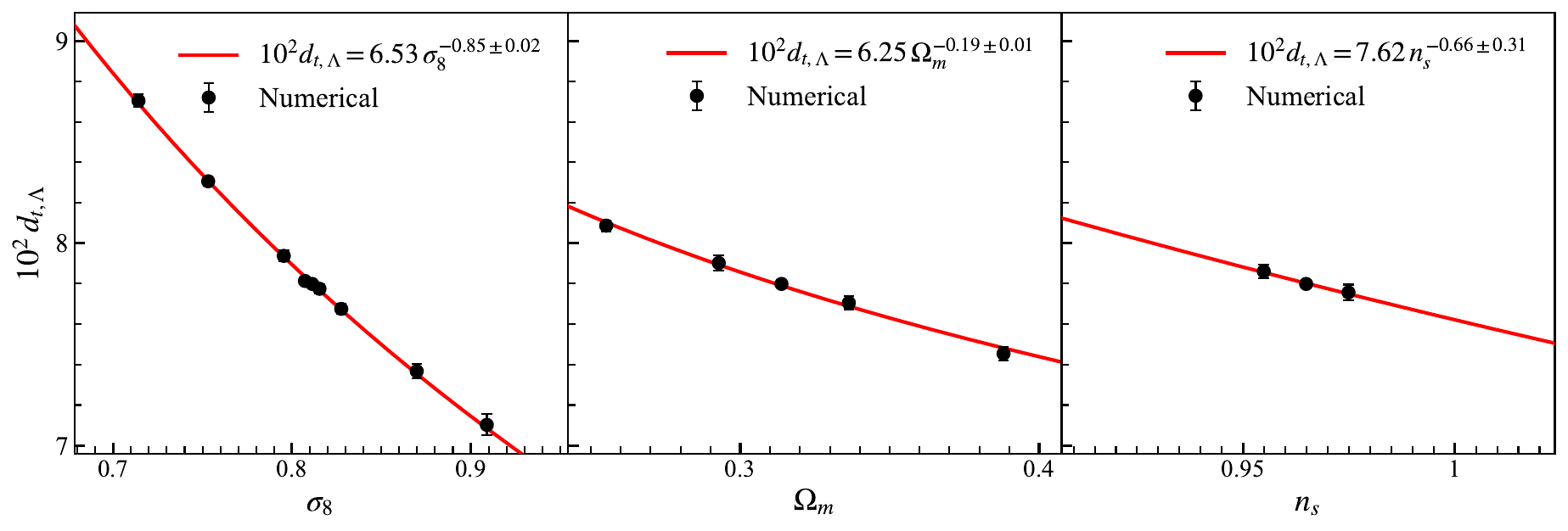}
\caption{\label{fig:dt_nonpl} 
Dependencies of the strengths of the perpendicular alignments of void-surface galactic halos on three key cosmological parameters of the $\Lambda$CDM universe (black circles), 
compared with the power-law model with best-fit power-law index (red lines).
The key cosmological parameters other than the one that shows a variation are set at the same values as the Planck $\Lambda$CDM universe.}
\end{figure}

To numerically prove this speculation, we investigate if and how $d_{t}$ changes with the variation of each of the three parameters, $\sigma_{8}$, $\Omega_{\Lambda}$, and $n_{s}$. 
First, to see the behavior of $d_{t}(\sigma_{8})$, we analyze eight additional AbacusSummit simulations performed for non-Planck $\Lambda$CDM cosmologies. 
The key cosmological parameters other than $\sigma_{8}$ are all set at the Planck values for these additional simulations (from the second to tenth rows in table~\ref{tab:lcdm}). 
For each case, we numerically determine $p(\cos\theta)$ and find the best-fit value of $d_{t}$ by fitting $p(\cos\theta)$ to eq.(\ref{eqn:pcost}). It turns out that even for these 
additional non-Planck $\Lambda$CDM cosmologies, eq.~(\ref{eqn:pcost}) excellently agrees with the numerical results. The left-panel of figure~\ref{fig:dt_nonpl} plots $d_{t}$ versus 
$\sigma_{8}$ (black circles), confirming that the perpendicular alignments of void-surface galactic halos indeed become stronger as $\sigma_{8}$ decreases, as speculated. 
Noting a power-law behavior of $d_{t}(\sigma_{8})$, we compare this numerical result to a power-law relation, $d_{t}=A_{d}\sigma^{n_{d}}_{8}$ and find the amplitude to be 
$A_{d}\approx 6.53$ and the power index to be $n_{d}\approx -0.85$ with the help of the $\chi^{2}$-minimization. The best-fit power-law relation 
is also shown as red solid curves in the left panel of figure~\ref{fig:dt_nonpl}. 
 
In a similar manner,  we also investigate how $d_{t}$ changes with each of $\Omega_{\Lambda}$ and $n_{s}$. In the AbacusSummit suite are available four (two)  
additional simulations, the initial conditions of which differ only in $\Omega_{\Lambda}$ ($n_{s}$) from the Planck $\Lambda$CDM case.  Repeating the whole process but 
with these additional simulations, we detect similar power-law behaviors of $d_{t}(\Omega_{\Lambda})$ and $d_{t}(n_{s})$ (red solid curves), and find their best-fit power-law indices 
and amplitudes, which are all displayed in the middle and right panels of figure~\ref{fig:dt_nonpl}, respectively.  The cosmologies with larger amounts of DE and lower spectral indices tend to 
yield stronger perpendicular alignments of void-surface galactic halos.  It is, however, worth noting that $d_{t}$ is the most sensitive to $\sigma_{8}$ among the three key cosmological 
parameters of a $\Lambda$CDM cosmology. These results support our claim that the perpendicular alignments of void-surface galactic halos are enhanced in a universe where the voids 
tend to form earlier. 

\subsection{Numerical results for the $w$CDM and $w_{0}w_{a}$CDM cases}\label{sec:wcdm}

Now that the strengths of the perpendicular alignments of void-surface galactic halos quantified by $d_{t}$ in eq.(\ref{eqn:pcost}) have been found to depend on the amount of DE, we would like to find 
an answer to a naturally arising question of whether or not $d_{t}$ also varies with the DE equation of state $w$. As stated in section~\ref{sec:intro}, we speculate that the answer to this question 
should be positive: if the perpendicular alignments of void-surface galactic halos were truly facilitated by the rapid acceleration of cosmic voids, they should be stronger (weaker) in a $w$CDM cosmology 
with $w<-1$ ($w>-1$ ) where the universe and its void structures accelerate more (less) rapidly than those in the $\Lambda$CDM case. 
To verify this speculation, we first  analyze four additional AbacusSummit simulations, each of which is performed for a $w$CDM cosmology with a different constant $w$ (see table~\ref{tab:wcdm}).
The initial conditions of these additional simulations were the same as the Planck values except for $w$~\cite{summit1}. 

Repeating the whole calculations described in section~\ref{sec:lcdm} but with these four additional AbacusSummit simulations, we determine the best-fit value of 
$d_{t}$ as well as $p(\cos\theta)$ for each of the four $w$CDM cosmologies.  
The middle panel of figure~\ref{fig:pcost} plots the same as its left panel but for one exemplary $w$CDM cosmology with $w=-0.9$, demonstrating the success of 
eq.(\ref{eqn:pcost}) in describing the behavior of the numerically obtained $p(\cos\theta)$ even for the $w$CDM case. 
Although we display the results only for one $w$CDM cosmology,  the validity of eq.(\ref{eqn:pcost}) is confirmed for all of the four $w$CDM cosmologies. 

\begin{figure}[tbp]
\centering 
\includegraphics[width=0.85\textwidth=0 380 0 200]{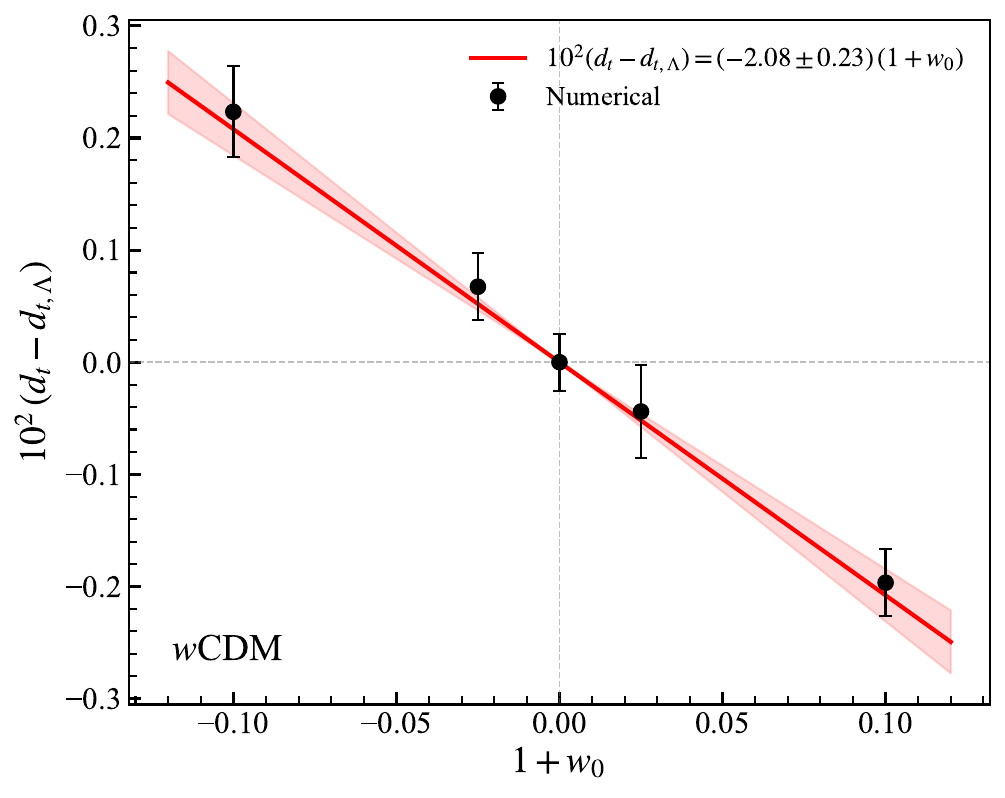}
\caption{\label{fig:dt_w} 
Numerical results for the differences in the strengths of the perpendicular alignments (black filled circles) between the $\Lambda$CDM and $w$CDM models 
as a function of the constant DE equation of state, $w_{0}$, compared with the bilinear model with best-fit slope, $\alpha$ (red solid lines).}
\end{figure}

Figure~\ref{fig:dt_w} plots $d_{t}-d_{t,\Lambda}$ versus $w$ (black circles), where $d_{t}$ is the best-fit value of this parameter for a given $w$CDM cosmology while 
$d_{t,\Lambda}$ is the best-fit value for the Planck $\Lambda$CDM case.  As can be seen, when $w<-1$ ($w> -1$), we have $d_{t}> d_{t,\Lambda}$ ($d_{t}<d_{t,\Lambda}$).
In other words, when the spacetime and voids undergo more rapid accelerations, the void-surface galactic halos develop stronger perpendicular alignments with the directions toward 
the void centers. Noting that the values of $\Delta d_{t}\equiv d_{t}-d_{t,\Lambda}$ exhibit an almost perfect linear decrement with the increment of $1+w$, 
we model the $(1+w)$-dependence of $\Delta d_{t}$ as a straight line of $\Delta d_{t}=\alpha(1+w)$ (red solid lines in figure~\ref{fig:dt_w}), the slope of which is determined via the 
$\chi^{2}$-statistics to be $\alpha\approx (-2.08\pm 0.23)\times 10^{-2}$. As can be seen, this straight-line model with the best-fit slope indeed matches the numerical results quite well. 

A crucial implication of results shown in figures~\ref{fig:dt_nonpl}-\ref{fig:dt_w} is the following:  it is not only the amount of exposure time to the accelerating expansion of voids but also 
the instantaneous rate of void expansions that affects the strengths of perpendicular alignments of void-surface galactic halos.  In a $w$CDM universe with $w<-1$, the voids must be exposed to the 
accelerating expansion for a shorter period of time than the case of $w\ge -1$. Nevertheless, their surface galactic halos tend to have stronger perpendicular alignments, as far as the other initial conditions, 
$\sigma_{8}$, $\Omega_{\rm de}$ and $n_{s}$ are fixed at the same values. Our explanation for this phenomenon is that more rapid accelerations of voids must drive more robustly and more coherently 
the DM particles at the void boundaries to move along tangential directions, resulting in stronger perpendicular alignments of void-surface galactic halos. Between the two quantities, the amount of 
exposure time to the accelerating expansion and the instantaneous rate of void expansions,  the former seems to make a larger contribution to enhancing the perpendicular alignments of void-surface 
galactic halos than the latter, given our finding of $d_{t,\Lambda}\gg \alpha(1+w)$ at a fixed value of $d_{t}$.  However, the value of $d_{t}$ scales {\it linearly} with $w$, 
while its scaling relations with each of $\sigma_{8}$, $\Omega_{\rm de}$ and $n_{s}$ have a milder slope less than unity, which indicates that the strengths of perpendicular alignments of void-surface 
galactic halos must be more sensitive to any departure of $w$ from the $\Lambda$CDM value of $-1$.

We analyze five additional AbacusSummit simulations performed for five different $w_{0}w_{a}$CDM cosmologies characterized by five different combinations of 
$w_{0}$ and $w_{a}$ (see table~\ref{tab:wcdm}). The other key cosmological parameters of these five $w_{0}w_{a}$CDM cosmologies were also set at the identical Planck values~\cite{summit1}. 
Following the same computational steps but with a dataset from these five additional simulations leads us to determine $p(\cos\theta)$ and the best-fit value of $d_{t}$ for each case of $(w_{0},w_{a})$. 
The right-panel of figure~\ref{fig:pcost} plots the same as its middle panel but for one exemplary $w_{0}w_{a}$CDM cosmology ($w_{0}=-0.7$ and $w_{a}=-0.4$), 
confirming that the analytic formula, eq.(\ref{eqn:pcost}), validly describes the numerically obtained $p(\cos\theta)$ even for the $w_{0}w_{a}$CDM case. 
In fact, good agreements between the analytic formula and the numerical results have been found for all of the five $w_{0}w_{a}$CDM cosmologies. 

\begin{figure}[tbp]
\centering 
\includegraphics[width=0.85\textwidth=0 380 0 200]{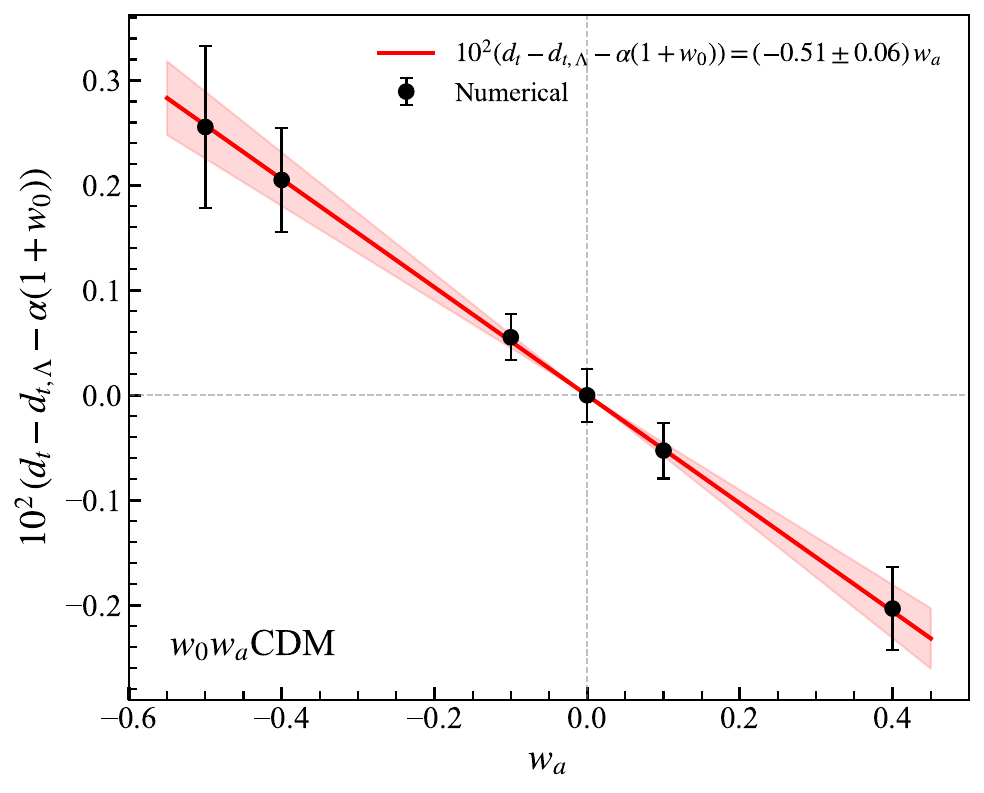}
\caption{\label{fig:dt_w0wa} 
Numerical results for the differences in the strengths of the perpendicular alignments (black filled circles) between the $\Lambda$CDM and $w_{0}w_{a}$CDM models 
as a function of the time-varying part of the DE equation of state, $w_{a}$, compared with the bilinear model with best-fit slope, $\beta$ (red solid lines).}
\end{figure}

Figure~\ref{fig:dt_w0wa} plots $\Delta d_{t}$ versus $w_{a}$, where $\Delta d_{t}$ is defined as 
$\Delta d_{t} \equiv d_{t}-d_{t,\Lambda}-\alpha(1+w_{0})$ with the same constant $\alpha$ found for the $w$CDM case. 
As can be seen,  the numerical results indicate $\Delta d_{t}>0$ ($\Delta d_{t}<0$) when $w_{a}<0 (w_{a}>0)$, which implies that 
if DE evolves more rapidly toward a more negative equation of state, the void-surface galactic halos develop stronger perpendicular alignments 
with the directions to the void centers.  This result can also be well accommodated by the aforementioned scenario that the higher instantaneous rate of accelerating void expansions can 
drive more coherent and rapid tangential motions of the DM particles at the void boundaries, resulting in stronger perpendicular alignments of void-surface galactic halos. 
Noting also that $\Delta d_{t}$ shows a linear scaling with $w_{a}$, we model it as a straight line of 
$\Delta d_{t}=\beta\,w_{a}$, too, and determine the best-fit value of $\beta$ to be $\beta\approx -(0.52\pm 0.03)\times 10^{-2}$ for which $\alpha$ is fixed at the constant value obtained 
for the $w$CDM case.  Overall, the following bilinear model turns out to effectively describe the $w(a)$-dependence of the deviation of $d_{t}$ from the $\Lambda$CDM counterpart, $d_{t,\Lambda}$: 
\begin{equation}
\label{eqn:bil}
d_{t}(w_{0},w_{a};\alpha,\beta)=d_{t,\Lambda}+\alpha(1+w_{0})+\beta\,w_{a}\, .
\end{equation}

\begin{figure}[tbp]
\centering 
\includegraphics[width=0.85\textwidth=0 380 0 200]{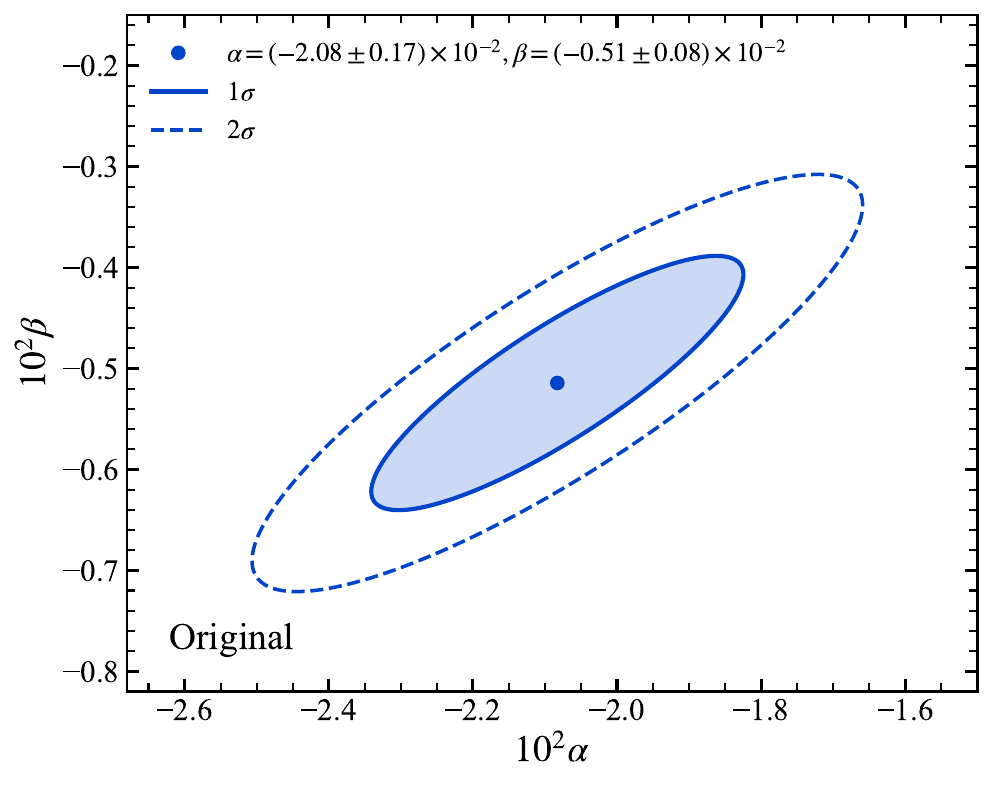}
\caption{\label{fig:alpha_beta} 
$1\sigma$ and $2\sigma$ contours of $\chi^{2}(\alpha,\beta)$ obtained by treating both of $\alpha$ and $\beta$ as adjustable parameters in 
eq.(\ref{eqn:bil}).}
\end{figure}

We also investigate if and how a possible mutual correlation between $\alpha$ and $\beta$ would alter their best-fit values and associated uncertainties as well.
In lieu of fixing $\alpha$ at the best-fit value obtained for the case of $w$CDM cosmologies, both of $\alpha$ and $\beta$ are treated as two adjustable parameters for this investigation. 
The marginalized errors as well as the best-fit values of $\alpha$ and $\beta$ are all redetermined by fitting eq.(\ref{eqn:bil}) to the numerical results via the $\chi^{2}$-statistics. 
Figure~\ref{fig:alpha_beta} plots the $68\%$ and $95\%$ contours of $\chi^{2}(\alpha,\beta)$, revealing the structure of covariance between the two parameters.  The linear correlation 
coefficient between $\alpha$ and $\beta$ is also determined to be $\approx 0.83$, indicating that they are almost linearly dependent on each other. 
As can be seen, although the marginalized errors become slightly larger than the un-marginalized ones, the redetermined best-fit values of $\alpha$ and $\beta$ 
are quite close to the original values obtained without considering their mutual correlation. This result supports our claim that the bilinear model, eq.~(\ref{eqn:bil}), with 
constant coefficients of $\alpha$ and $\beta$ are universally valid for all of the three classes of the background cosmologies. 

\begin{table*}[tbp]
\centering
\begin{tabular}{lccccccc}
\hline
\hline
\rule{0pt}{3.5ex}\noindent
Model & $w_0$ & $w_a$  & $N_{\rm void}$ & $N_{\rm vs}$ & $N_{\rm vs, m}$ & $N_{\rm vs, S}$ & $10^2 d_t$ \\
\hline
\rule{0pt}{3.5ex}\noindent
  c000 & $-1.0$ & $0.0$  & 634051 & 41432782 & 40272159 & 40912371 & $7.80 \pm 0.02$\\
\hline
\rule{0pt}{3.5ex}\noindent
  c108 & $-0.9$ & $0.0$  & 612012 & 41922146 & 40366762 & 40546311 & $7.60 \pm 0.02$\\
  c109 & $-1.1$ & $0.0$  & 653378 & 40612477 & 40272159 & 40532704 & $8.02 \pm 0.04$\\
  c121 & $-0.975$ & $0.0$  & 628454 & 41569968 & 40844814 & 40994402 & $7.75 \pm 0.04$\\
  c122 & $-1.025$ & $0.0$  & 640052 & 41226908 & 40272159 & 40931437 & $7.87 \pm 0.02$\\
\hline
\rule{0pt}{3.5ex}\noindent
  c002 & $-0.7$ & $-0.5$  & 596848 & 42083663 & 40274194 & 40532704 & $7.43 \pm 0.03$\\
  c110 & $-1.1$ & $0.4$  & 632243 & 41484995 & 40643097 & 40963987 & $7.80 \pm 0.03$\\
  c111 & $-0.9$ & $-0.4$  & 635321 & 41374846 & 40272159 & 40881219 & $7.80 \pm 0.04$ \\
  c123 & $-1.025$ & $0.1$ & 633871 & 41429191 & 40320835 & 40964072  & $7.80 \pm 0.02$ \\
  c124 & $-0.975$ & $-0.1$  & 634974 & 41370469 & 40272159 & 40898499 & $7.80 \pm 0.01$\\
\hline
\rule{0pt}{3.5ex}\noindent
\end{tabular}
\caption{\label{tab:wcdm}
Appellation of the background $w$CDM and $w_{0}w_{a}$CDM cosmologies, DE equation of state,  numbers of voids and void-surface halos belonging to the original ($N_{\rm vs}$), 
mass-controlled and sphericity-controlled samples, best-fit correlation parameter}
\end{table*}
Now that we have detected the effect of dynamical nature of DE on the strengths of the perpendicular alignments of void-galactic halos, we also would like to examine whether it is truly a {\it net} 
effect or not.  As shown in section~\ref{sec:lcdm}, the void-surface galactic halos in different ranges of $m_{t}$ and $S$ yield different values of $d_{t}$. Meanwhile, different cosmologies 
with different $w$ may well yield different $m_{t}$ and $S$ distributions of void-surface galactic halos. 
In other words, the observed signal of the $w(a)$-dependence of $d_{t}$ could be at least partially contributed by any differences in the $m_{t}$ and $S$ distributions 
among the different DE cosmologies. Therefore, to isolate the net effect of the dynamic nature of DE on the strengths of the perpendicular alignments of void-surface galactic halos, 
it should be desirable to use the controlled samples of void-surface galactic halos that have no difference in the $m_{t}$ and $S$ distributions among different DE cosmologies. 

\begin{figure}[tbp]
\centering 
\includegraphics[width=0.85\textwidth=0 380 0 200]{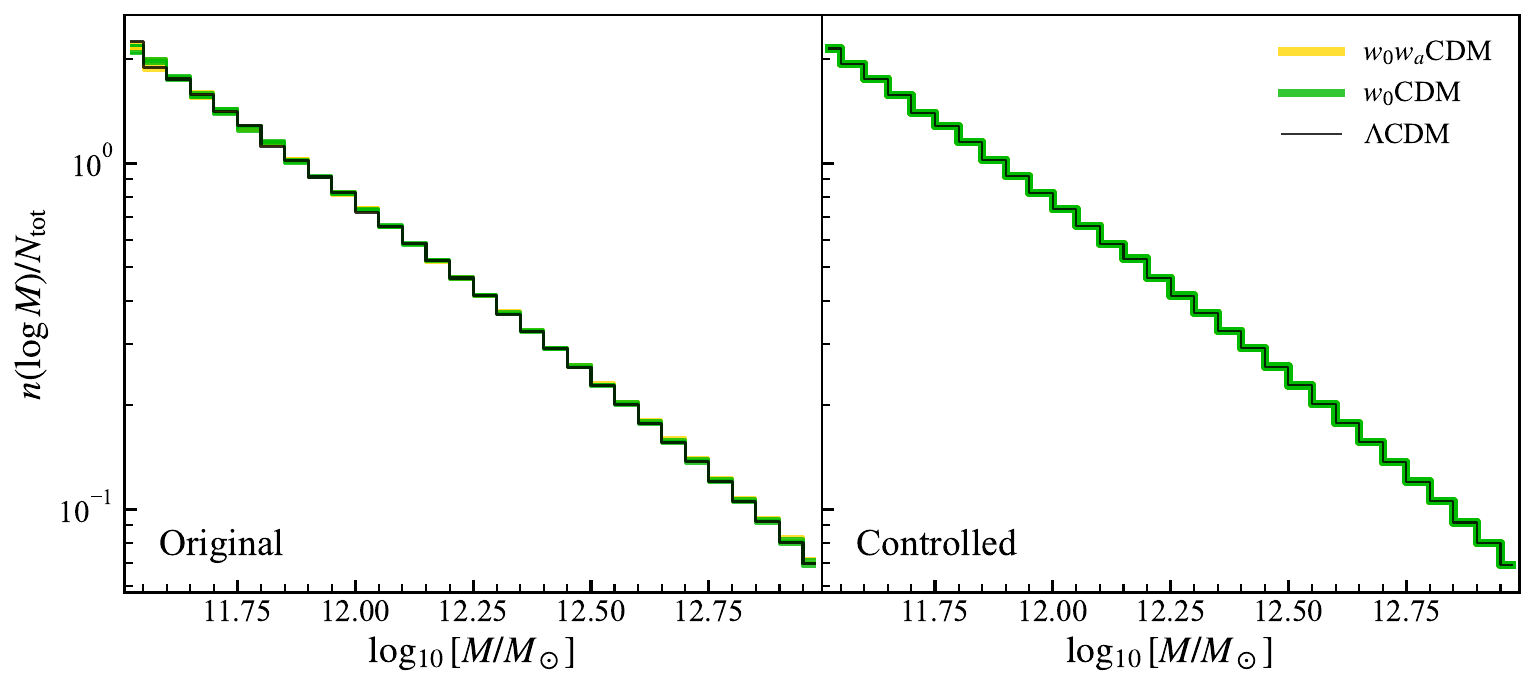}
\caption{\label{fig:m_cont} 
Fractional number counts of the void-surface galactic halos versus their logarithmic masses, $m_{t}$, from the original and $m$-controlled samples 
for the one $\Lambda$CDM, four $w$CDM and five $w_{0}w_{a}$CDM cosmologies.}
\end{figure}

The left-panel of figure~\ref{fig:m_cont} plots the fractional number distribution of void-surface galactic halos, $n_{\rm vs}(m_{t})/N_{\rm vs}$, as a function of $m$ from the ten different 
cosmologies (one $\Lambda$CDM, four $w$CDM and five $w_{0}w_{a}$CDM), where $n_{\rm vs}(m)$ is the number of void-surface galactic halos in a differential interval 
of $[m_{t},m_{t}+dm_{t}]$. As can be seen, the ten different DE cosmologies indeed differ in  $n_{\rm vs}(m_{t})/N_{\rm vs}$ from one another. To eliminate any spurious signal of the 
$w(a)$-dependence of $d_{t}$ caused by these differences, we create a controlled sample of void-surface galactic halos for each of the ten cosmologies: splitting the range of 
$m_{t}$ into multiple short intervals, we deliberately cull a specific number of void-surface galactic halos at each $m$-bin, which makes all of the ten cosmologies have identical 
distributions of $n_{\rm vs}(m_{t})/N_{\rm vs}$. The fractional number counts from these $m_{t}$-controlled samples of the void galactic halos are shown in the right panel of 
figure~\ref{fig:m_cont}. 

\begin{figure}[tbp]
\centering 
\includegraphics[width=0.85\textwidth=0 380 0 200]{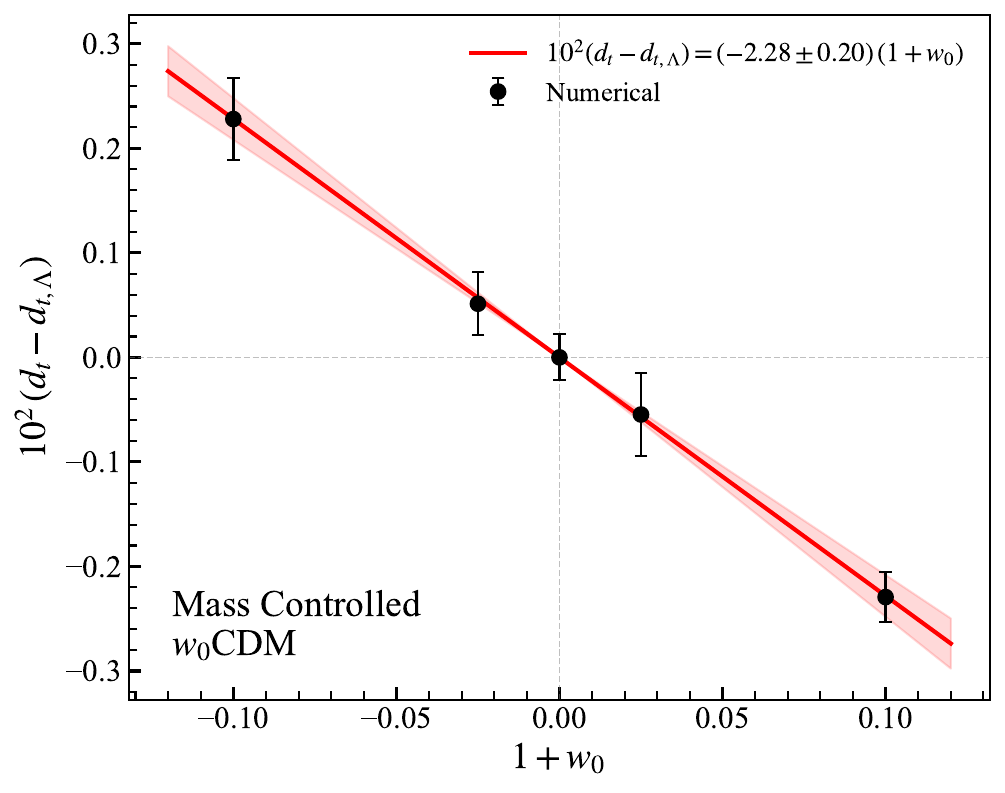}
\caption{\label{fig:dt_w_mcont} 
Same as figure~\ref{fig:dt_w} but from the $m_{t}$-controlled samples.}
\end{figure}
\begin{figure}[tbp]
\centering 
\includegraphics[width=0.85\textwidth=0 380 0 200]{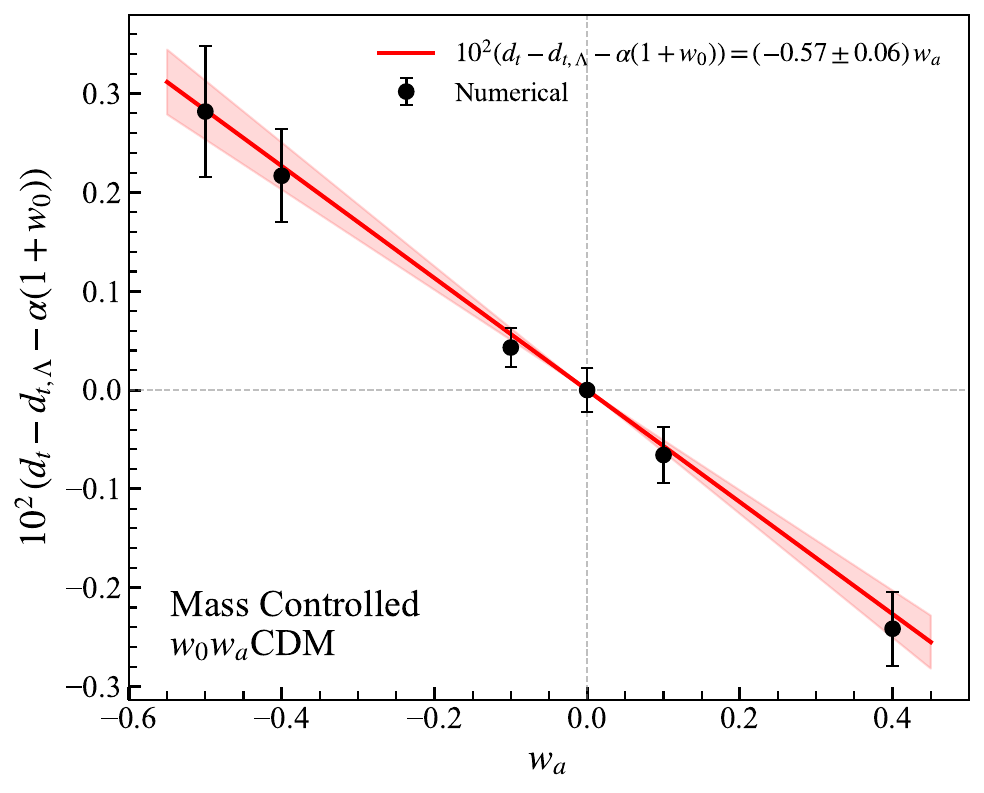}
\caption{\label{fig:dt_w0wa_mcont} 
Same as figure~\ref{fig:dt_w0wa} but from the $m_{t}$-controlled samples.}
\end{figure}
Repeating the whole procedure but with the $m_{t}$-controlled sample of void-surface galactic halos for each of the $w$CDM and $w_{0}w_{a}$CDM cosmologies, 
we redetermine $d_{t}$, its deviation from the $\Lambda$CDM case, $\Delta d_{t}$, and the coefficients, ($\alpha,\beta)$, which are all displayed in figures~\ref{fig:dt_w_mcont}-\ref{fig:dt_w0wa_mcont}. 
As can be seen, the results from the $m_{t}$-controlled samples are quite similar to those from the original samples of void-surface galactic halos, 
in spite of no difference in $m_{t}$-distributions among the ten DE cosmologies.  The bilinear model, eq.(\ref{eqn:bil}), for $\Delta d_{t}$ still holds true, confirming its robustness.  
Note, however, that the $m_{t}$-controlled samples of void-surface galactic halos yield slightly higher values of $\vert\alpha\vert$ and $\vert\beta\vert$, which indicates that when the spurious 
signals caused by the differences in $m_{t}$-distributions among different DE cosmologies are eliminated, the strengths of the perpendicular alignments of void-surface galactic halos become 
even more sensitive to the variation of the DE equation of state. 

\begin{figure}[tbp]
\centering 
\includegraphics[width=0.85\textwidth=0 380 0 200]{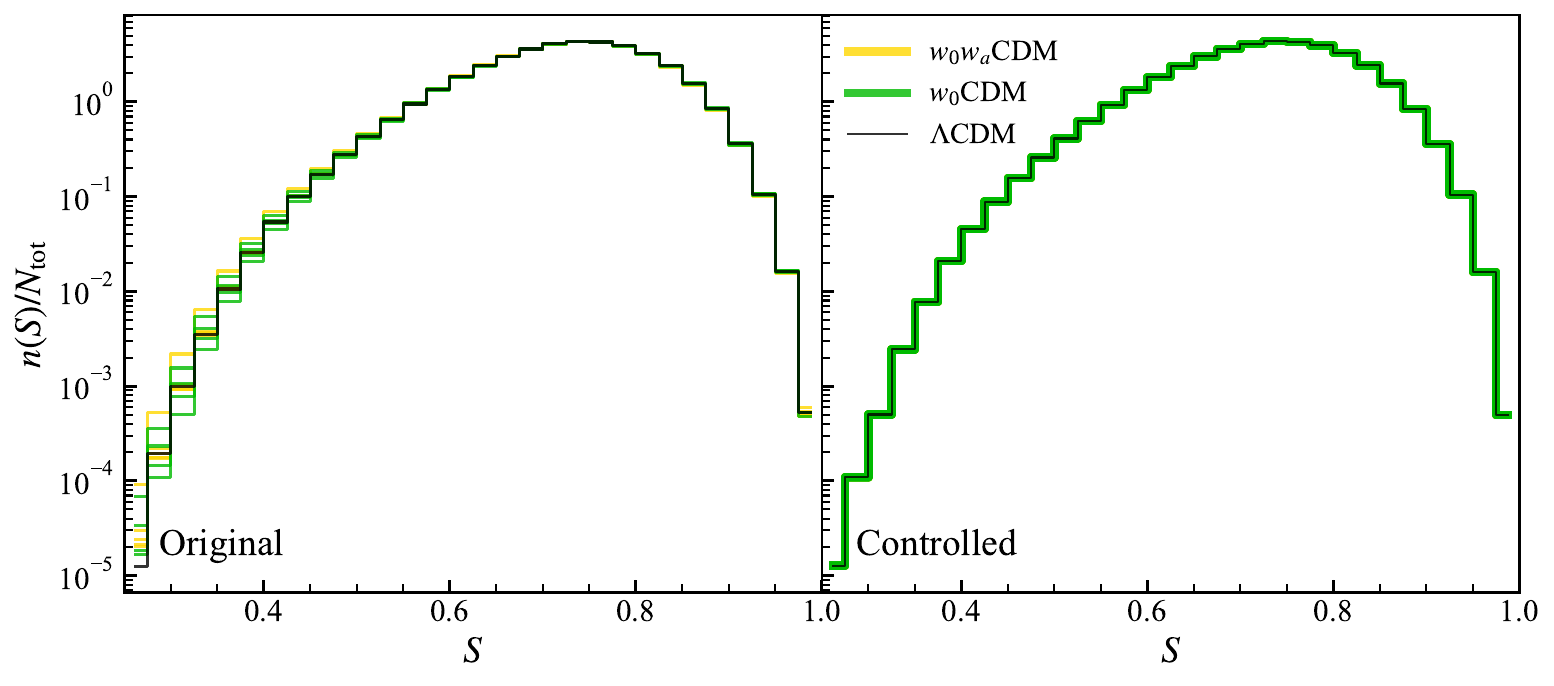}
\caption{\label{fig:s_cont} 
Number fractions of the void-surface galaxies versus their sphericities, $S$, from the original and controlled samples for the ten different cosmologies.}
\end{figure}

\begin{figure}[tbp]
\centering 
\includegraphics[width=0.85\textwidth=0 380 0 200]{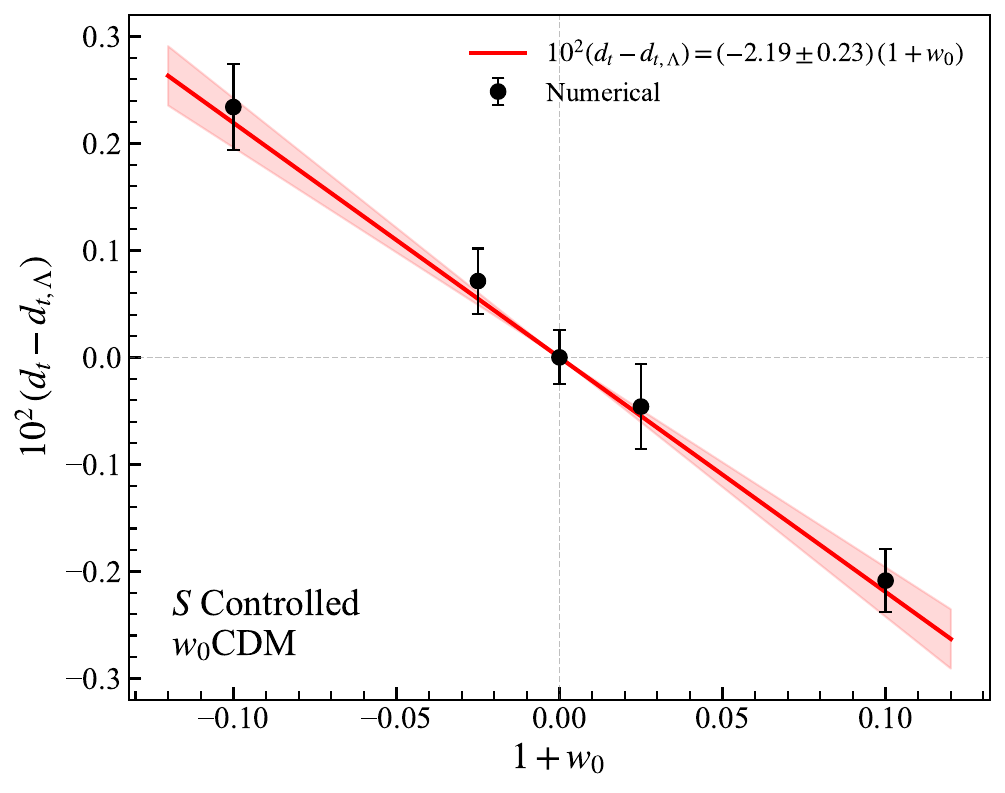}
\caption{\label{fig:dt_w_scont} 
Same as figure~\ref{fig:dt_w} but from the $S$-controlled samples.}
\end{figure}

\begin{figure}[tbp]
\centering 
\includegraphics[width=0.85\textwidth=0 380 0 200]{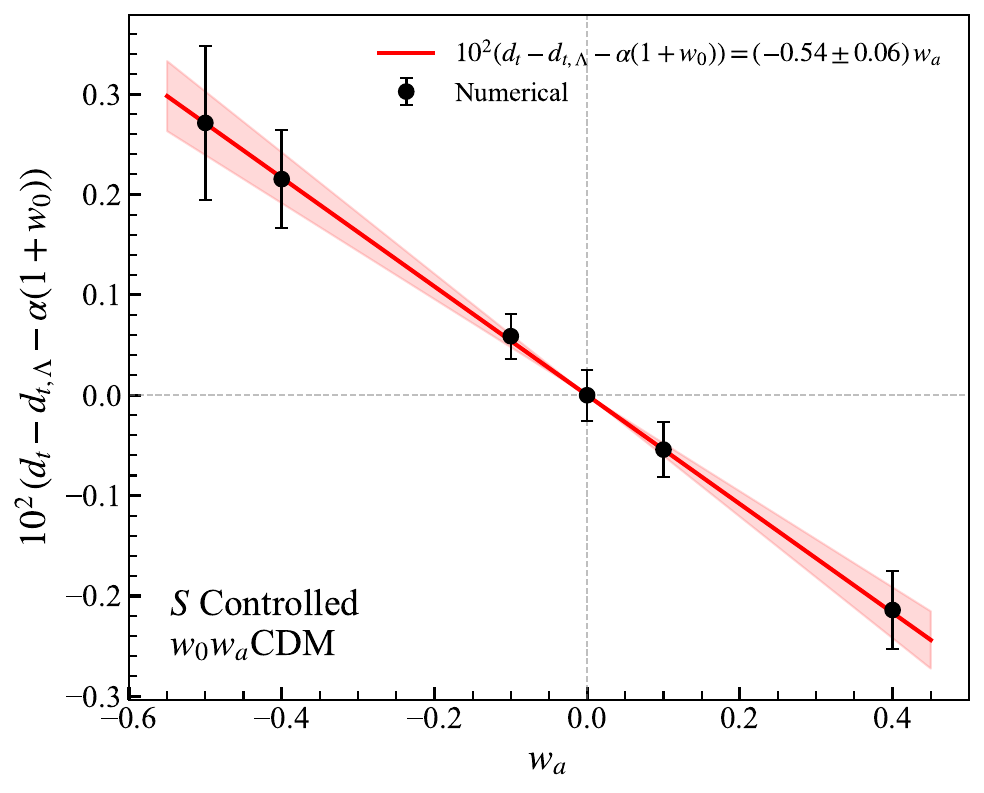}
\caption{\label{fig:dt_w0wa_scont} 
Same as figure~\ref{fig:dt_w0wa} but from the $S$-controlled samples.}
\end{figure}

In a similar manner, we also create the $S$-controlled samples of void-surface galactic halos to remove any possible spurious signals induced by the differences in $S$-distributions 
among the ten backgrounds. Figure~\ref{fig:s_cont} shows the fractional number counts from the original and $S$-controlled samples of void galactic halos from the ten DE cosmologies in the left 
and right panels, respectively. 
The behaviors of $\Delta d_{t}[w(a)]$ and the coefficient values of power-law relations from these $S$-controlled samples are all shown in figures~\ref{fig:dt_w_scont}-\ref{fig:dt_w0wa_scont}, which 
reveal that when the $S$-controlled samples of void-surface galactic halos are used, $\Delta d_{t}[w(a)]$ is approximated by a steeper straight line. 
The results shown in figures~\ref{fig:dt_w_mcont}-\ref{fig:dt_w0wa_scont} point to the success of our bilinear model in describing the effect of the dynamic DE on the degree of the departure of $d_{t}$ 
from the $\Lambda$CDM case. In the last three columns of table~\ref{tab:wcdm} are listed the numbers of void-surface galactic halos and the best-fit values of $d_{t}$ from the $m_{t}$ and $S$-controlled 
samples for each cosmology. 

\section{Discussion and conclusion}\label{sec:con}

The most urgent mission in cosmology is to constrain the DE equation of state, $w$, as tightly as possible, and to investigate if it evolves with time.  
The tightest constraint on $w$ will bring light not only on the true nature of DE but also on the early universe physics as well as on the final destination of our universe. 
This mission has so far been primarily carried out by employing the linear statistics such as the cosmic microwave background radiation (CMBR) temperature spectrum, baryonic 
acoustic oscillations (BAO) and luminosity-distance relation of type Ia supernovae (SNIa)~\cite{planck18,desi,DR04,jon-etal22}. 
The powers of these linear statistics lie in the fact that their dependencies on $w$ were mathematically derived from the linear perturbation theory without requiring any extra parameters 
(other than the key cosmological parameters including $w$). On one hand, it is the most advantageous merit of the linear statistics that the first principles dictate their $w$-dependencies. 

On the other hand, it implies that their $w$-dependencies are rather indirect and thus not so sensitive to the variation of $w$. 
Besides, if the cosmic acceleration was caused not by $\Lambda$ but by dynamical DE characterized by evolving equation of state, $w(a)$, then the fundamental principles alone 
fail to trace $w(a)$, due to the lack of established physics for it.  To complement the linear statistics, much effort has been made to develop non-linear statistics that are to achieve 
higher $w$-sensitivity at the cost of introducing extra free parameters~\cite{MW03,guz-etal08,boy-etal21}.  
For example, the nonlinear redshift space distortion effect has been demonstrated to be a powerful nonlinear probe of $w(a)$~\cite{guz-etal08}, for which the pairwise velocity 
dispersion was introduced as an extra parameter. 

A desirable non-linear statistics as a probe of DE should be the one that can be describable by a simple analytic formula with the least number of extra free 
parameters expressed in explicit and direct terms of $w$. In the current work, we have numerically proven that the perpendicular alignments of void-surface galaxies should 
in principle be such an idealistic nonlinear-statistics: it is well approximated by a single parameter formula, which is valid regardless of the DE equation of state;  the difference in 
the best-fit value of the single parameter between the $\Lambda$CDM and $w$CDM cosmologies linearly scales with both of $w_{0}$ and $w_{a}$~\cite{CP01,linder03}. 
Another upside of our new non-linear statistics is that it does not require high-$z$ data, in contrast to the majority of the conventional nonlinear statistics that depend implicitly on 
$w$ through their dependencies on the density growth rate, for the accurate measurements of which usually require high-$z$ data ($z\gg 0.1$)~\cite{ZL26}.

Yet, there is a couple of challenging issues that must be addressed prior to a practical application of this new nonlinear statistics to probe the DE equation of state. 
First of all, it is not guaranteed that the same (or at least similar) bilinear relation would exist between $d_{t}$ and $w$ if the shape axes are determined 
from the stellar components. Given that in real observations what can be measured is the two-dimensional shape axes of elliptical galaxies rather than the three dimensional counterparts 
of triaxial DM halos on void surfaces and that the shapes of DM halos are not perfectly aligned with those of visible galaxies at virial radii~\cite{vel-etal15}, 
it will be essentially necessary to use hydrodynamic simulations performed for various $w$CDM cosmologies and to inspect non-parametrically whether or not the same bilinear model holds true 
even when $d_{t}$ is determined from the two dimensional shape axes of visible stellar components. 
The second issue is that the other initial conditions than the ones considered in the current work might as well affect the strengths of the perpendicular alignments of void-surface galactic halos.
Given our finding that the value of $d_{t}$ depends most sensitively on the amplitude of the initial density power spectrum, it might also depend on the running spectral index as well as on the neutrino mass. 
To verify the power of this new nonlinear statistics based on the strengths of perpendicular alignments of void-surface galactic halos, it will be definitely necessary to explore how it behaves 
in a larger parameter space and to examine any possible degeneracy among the initial conditions in its values. Our future work will be in the direction of addressing these critical issues and 
undertaking an observational application of this new nonlinear statistics to probe the dynamic nature of DE.

\begin{acknowledgments}

JL acknowledges the supports by Basic Science Research Program through the National Research Foundation (NRF) of Korea funded by the Ministry of Education (RS-2025-00512997). 
GK acknowledges that this research was supported by Basic Science Research Program through the National Research Foundation of Korea (NRF) funded by the Ministry of Education (RS-2025-25408975)

\end{acknowledgments}

\end{document}